*Article*

# Exploratory Data Analysis of Electric Tricycle as Sustainable Public Transport Mode in General Santos City Using Logistic Regression

G. L. Cueto [1,a], F.A.A. Uy [2,b], and K.A. Diaz[3,c,*]

1,2,3 School of Graduate Studies
Mapúa University, Muralla St., Intramuros, Manila 1002, Philippines
E-mail: [a]geoffrey_cueto@yahoo.com, [b]drfrancisuy@gmail.com, [c,*]keithdiaz0611@gmail.com(Corresponding author)

**Abstract.** General Santos City, as the tuna capital of the Philippines, relies with the presence of tricycles in moving people and goods. Considered as a highly-urbanized city, General Santos City serves as vital link of the entire SOCKSARGEN region's economic activities. With the current thrust of the city in providing a sustainable transport service, several options were identified to adopt in the entire city, that includes cleaner and better transport mode. Electric tricycle is an after sought alternative that offers better choice in terms of identified factors of sustainable transport: reliability, safety, comfort, environment, affordability, and facility. A literature review was conducted to provide a comparison of cost and emission between a motorized tricycle and an e-tricycle. The study identified the existing tricycle industry of the city and reviewed the modal share with the city's travel pattern. Actual roadside and onboard survey were conducted both to passengers and drivers from the identified terminal to their destination. The survey revealed a number of hazards were with the current motorized tricycle that needs to address for the welfare of the passengers and drivers. With the application of exploratory data analysis using Python program, the study favors the shift to adopting E-tricycle. The model derived from binary logistics regression provided a 72.72% model accuracy. Based from the results and findings, electric tricycle can be an alternative mode of public transport in the city that highly support sustainable option that provides local populace to improve their quality of life through mobility and economic activity. Further recommendation to local policy makers in the transport sector of the city include the clustering of barangays for better traffic management and franchise regulation, the inclusion of transport-related infrastructure related to tricycle service with their investment planning and programming, the roll out and implementation of tricycle code of the city, and the piloting activity of introducing e-tricycle in the city.

**Keywords:** E-tricycle, exploratory data analysis, python, logistic regression, sustainable public transport

# 1. Introduction

## 1.1 Background of the Study

The use of tricycle has emerged to be the mode of transportation in most developing cities of the world. Tricycle (motorized and non-motorized) is the prime mode of intra-city transportation though considered as informal sector, its impact on transportation and management is beyond question. In China, working tricycles continue to provide social and economic needs despite the threat of being phased out due to the neoliberal mobility developing within the city. In India, the same mode of transportation is used by common people which posed a major challenge to the growing population on how the government would maintain good air quality. In Nigeria, the use of motorized tricycle is encouraged within the city limits to decongest the traffic. In Metro Manila, Philippines, it is the prevailing mode of transportation using secondary streets occupying the front seat in local policy making in local governance and urban development. In General Santos City Philippines, there is a high utilization of public transport. The modal share of the city revealed that 81% of school and work-related trips use various mode of public transport. On the average, it is 374,000 person-trips were produced, of which 19% use private transport and 81% is public transport. Of the public mode of transport, tricycles serve the highest number of person-trips at 42%. The trend on the modal share continue to decline on jeepneys and multicab with the tremendous increase of tricycles in the city.

In the Philippines, the emergence of electric tricycle or E-trike is much more comparative with the same as tricycle in terms of configuration or almost as similar to that of the traditional rickshaw which is three-wheeled, and can ferry several passengers and dependent on the body design and capacity of the motor engine. E-trike is run by electricity while the latter is gas-fed. A typical design popular in the Philippines is that a sidecar which is attached to the side of a motorcycle for carrying passengers.

The use of motorized tricycle is already a part of every residents of General Santos City with the everyday living. It is used as a mode of commuting, freight delivery system, private family service and source of income. Based from the Local Public Transport Route Plan of the city, the significant share of public transport trips emphasizes the role of public transport in the city. Further, the city considers tricycles as major transport mode catering to the trips generated by people according to various purposes.

The use of tricycle as a mode of transportation actually poses hazards and adverse effects on the environment's air quality. Further, about 34% of total vehicular population in the Philippines is made up of two- and three-wheelers which also become the major contributor to pollution and other environmental and health hazards. The use of motorized tricycle is also identified as the source of noise and air pollution exposing the public to a greater danger of health-related problems. For instance, a 2-stroke engine tricycle is known to contribute to air pollution in Jakarta Indonesia. Tricycles are identified as one of the public transport systems causing excessive gas emission contributing to lead and nitrogen dioxide release in the atmosphere.

## 1.2 Statement of the Problem

The emergence of tricycles as the major public transport mode in General Santos City is somehow add to the growing economy of informal transport sector. As supported by good infrastructure of the city, tricycle growth in the city considered as phenomenon that edge higher capacity public transport. With the increase of motorized tricycles in the city, pose several problems that affects the environment and the public health.

This study is conducted in order to determine the sustainability of providing e-tricycle as an alternative mode of transport in General Santos City based from the perception and acceptability of passengers and drivers. Moreover, it underscores possible policy recommendation for the city's current transportation management plan.

## 1.3 Objective of the Study

The main objective of conducting this study is to determine the sustainability of adopting e-tricycle as an alternative mode of transport in General Santos City.
The specific objectives are:
- To characterize the existing tricycle industry in General Santos City.
- To describe the hazards of the existing tricycle service to the environment and the community.
- To assess the possible transport policy implication of adopting e-tricycle in the city.

## 1.4 Scope and Limitations

The scope of the study will examine the existing tricycle services of General Santos City. It provided a comparison between the existing tricycle in the city and the proposed e-tricycle in terms of return of investment, and contribution of their emission to air quality and sustainability. The study was limited to the tricycle service in the geographical setting of General Santos City.

## 1.5 Significance of the Research

This research study was able to provide General Santos City to adopt a more sustainable mode of transport by virtue of e-tricycle. The study considers to be an input for policy maker to shift to a cleaner and environment-friendly tricycle services and at the same time create a platform of economic activity to local communities engaged in tricycle service.

# 2. Methodology

## 2.1 Framework of the Study

The purpose of the theoretical framework is to provide clearer picture on the scope of the study that will guide data collection. The variables present in this study such as tricycle services, routes and fare of tricycle, engine type, and modal share.

The relationships had been established with the application of exploratory data analysis using Python programming language. The model derived with the defined predictors of shifting to e-tricycle was established that served as a reference in deriving policy recommendation to General Santos City of possible adoption of e-tricycle as a sustainable mode of public transport in the city. The paradigm of Three E's of Sustainable Transport (CST,1997, Hall 2002) was applied to create sustainability of e-tricycle and consequential to policy recommendation regarding public transport in the city.

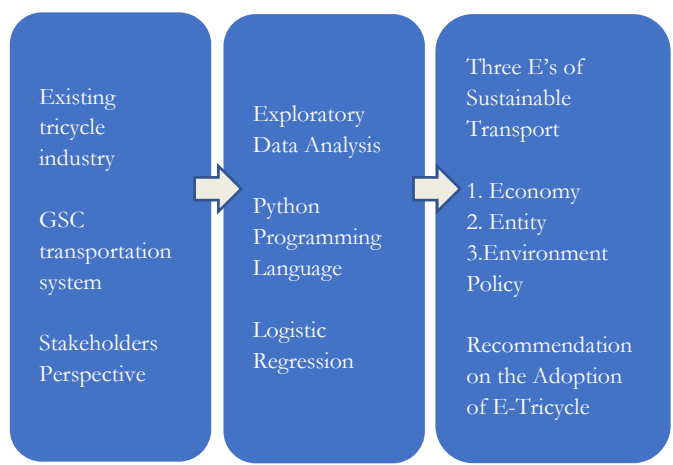

Fig. 1 Framework of the Study

**2.2 Research Design**

The study conducted roadside and onboard survey to the passengers and drivers residing or staying in General Santos City aimed to determine their travelling experience using the current motorized tricycle. The questionnaire design is a modified version based from the paper entitled "An Assessment of the Pilot Implementation of the Tricycle Modernization Program of the City Government of Imus, Cavite". The questionnaire is divided in to three (3) parts: 1. Passenger Profile 2. Tricycle Travelling Pattern and 3. Passenger Satisfaction.

In the Passenger Profile, the following were asked:
1. Gender
2. Age
3. Marital Status
4. Address
5. Occupation
6. Disability
7. Average Monthly Personal Income

In the Tricycle Travelling Pattern part, it was investigated the respondents' origin-destination, trip frequency, trip route, trip purpose, waiting time, cost of trip and travel time.

In the third part of the survey, respondents were asked to provide their ratings with regards to their riding experience based on the following aspects:
1. Reliability – commuter's waiting time, availability of tricycle units and flexibility of route;
2. Comfort and Convenience – seat comfort, cleanliness, travel time, ease of transfer, comfort while waiting;
3. Safety and Security – travel safety and security, risk involved and complaint system;
4. Affordability – value for trip;
5. Environment and Health Friendliness – smoke emission and noise;
6. Facility – presence of terminal, waiting station, dispatching system.

Moreover, the six (6) aspects were utilized as basis for additional fare commuters will add to the current fare for better travelling experience.

Data collected were analyzed using descriptive method of research for the demographics and travel pattern. More so, the data from the factors of travelling experience in the survey applied exploratory data analysis in Python which derived probability model of adopting electric tricycle shift.

The model used Environment, Safety, Reliability, Comfort, Facility and Affordability as predictors.

In Logistic Regression, the logits are assumed to be a linear function of the predictors,

$$L = \log\left(\frac{P(Y=1)}{1-P(Y=1)}\right) = \beta_0 + \sum_{i=1}^{n} \beta_i X_i \quad (1)$$

To solve for P(Y=1),

$$p = \frac{e^L}{1-e^L} \quad (2)$$

The parameters $\beta_i$ can be derived by Maximum Likelihood Estimation (MLE). The likelihood for a given observation $Y_j$ is;

$$lkl = \prod_{j=1}^{m} p^{y_j}(1-p)^{1-y_j} \quad (3)$$

The survey questionnaire developed for the drivers as respondent is composed of three (3) parts: 1. Driver's Profile; 2. Driving Pattern; and 3. Driving Experience

In the Driver's Profile, the following were asked
1. Gender
2. Age
3. Marital Status
4. Address
5. Occupation
6. Disability
7. Average Monthly Personal Income

The second part of the survey questionnaire determined the driver's knowledge about electric tricycle, willingness to shift to e-tricycle, and the time for the respondent's shift to e-tricycle.

The third part asked the respondents on:

1. Identified problems to be encountered in shifting to e-tricycle;
2. Needs to shift in using e-tricycle;
3. Perceived benefits of using e-tricycle; and
4. Hazards of using existing motorized tricycle.

The researcher also conducted an interview with the city officials of General Santos City to identify the programs, activities and projects related to transportation as well as their plans and strategies to achieve sustainable transport for the entire city. Moreover, numbers of emission testing centers were also visited and obtained copies of emission testing results of motorized tricycles.

## 3. Results and Discussion

| A. Socio-demographics attributes | |
|---|---|
| Sex | |
| M | 51.15% |
| F | 48.85% |
| Age | |
| < 16 | 25.95% |
| 16-21 | 32.06% |
| 22-35 | 21.37% |
| 36-50 | 12.21% |
| 51-60 | 6.11% |
| > 60 | 2.29% |
| Occupation | |
| Students | 31.30% |
| Private Employee | 19.85% |
| Government Employee | 6.11% |
| Entrepreneur/ Business | 11.45% |
| Vendor | 13.74% |
| Fisherman/ Farmer | 14.50% |
| Average Monthly Personal Income | |
| < 10,000 | 40.46% |
| 10,001 – 15,000 | 32.06% |
| 15,001 – 20,000 | 18.32% |
| 20,001 – 25,000 | 3.05% |
| 25,001 – 30,000 | 6.11% |
| >30,000 | 0.00% |
| Trip Commuting Pattern | |
| Frequency of tricycle use | |
| 1-2 times a day | 22.90% |
| 2-3 times a day | 66.41% |
| 3-4 times a day | 9.92% |
| >4 times a day | 0.76% |
| Trip purpose on using tricycle | |
| Home | 7.63% |
| Work | 45.80% |
| Social visit/shopping | 12.98% |
| School | 29.77% |
| Leisure/ recreational | 0.00% |
| Medical/personal | 3.82% |
| Others | 0.00% |
| Average travelling time going to tricycle terminal = 13.94 minutes | |
| Standard Deviation = 11.66 | |
| Average waiting time for tricycle = 12.08 minutes | |
| Standard deviation = 6.32 | |
| Average Travelling time using tricycles = 12.51 minutes | |
| Standard deviation = 4.78 | |
| Average travelling expense using tricycle (two way) = PHP 26.21 | |
| Standard deviation =. 15.06 | |

Table 1. Commuters Demographics and Trip Commuting Pattern

### 3.1 Passenger Satisfaction

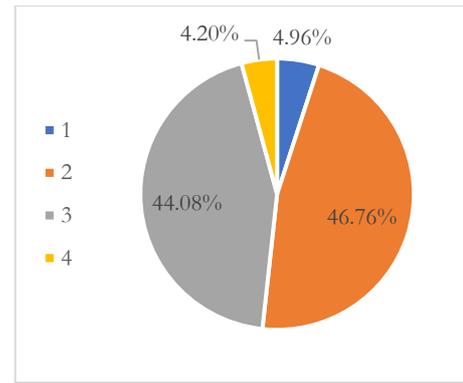

Fig. 2. Passenger Satisfaction on Reliability

Figure 2 shows that 44.08% of the commuters are satisfied with the reliability of the current tricycle system in General Santos City. On the other hand, 46.76% were dissatisfied. Commuter's dissatisfaction comes from the commuter's waiting time, contingency for vehicle breakdown, availability, and flexibility in route.

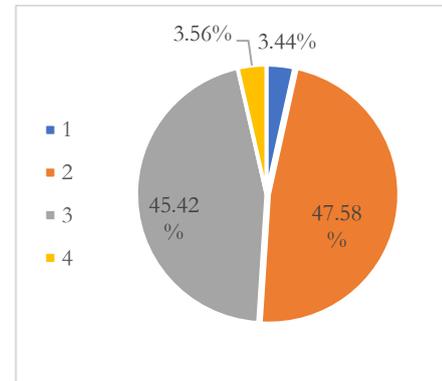

Fig. 3. Passenger Satisfaction on Comfort and Convenience

Figure 3 provides that 45.42% of the commuters are satisfied with the comfort and convenience the traditional tricycles can provide. Since the tricycle system is a point-to-point basis, most of the commuters find this very convenient while being comfortably sitting through the traffic the mode is traversing to.

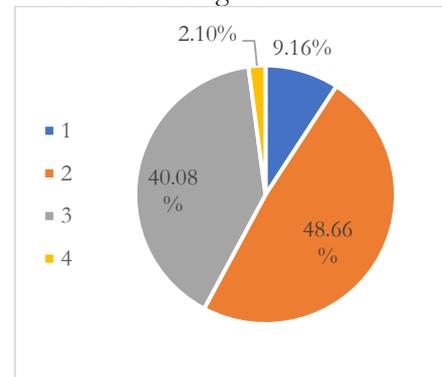

Fig. 4. Passenger Satisfaction on Safety and Security

In Figure 4, nearly half of the respondents (48.66%) find the current tricycle system dissatisfied. Commuters usually feels unsafe; this is mostly due to tricycles being an open vehicle. Others also considered the driver's driving habit which adds anxiousness to some commuters.

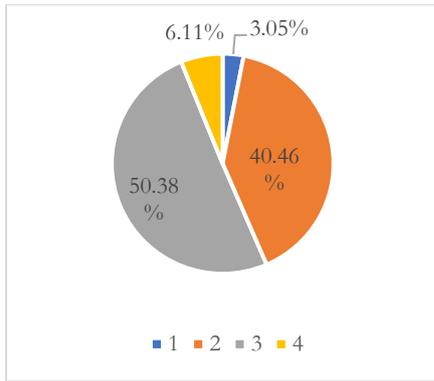

Fig. 5. Passenger Satisfaction on Affordability

Figure 5 shows that 56.49% (combined percentage of those who said "very satisfied", 6.11% and "Satisfied", 50.38%) of the commuters find the current tricycle system affordable. Commuters benefit from the affordability of the mode for a short distance travel along the area.

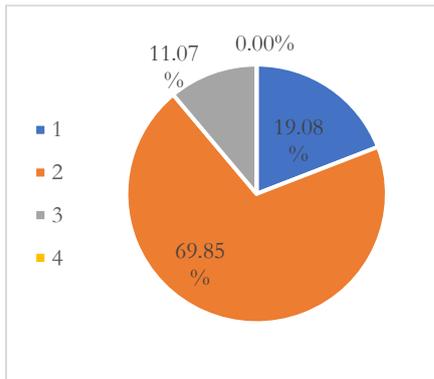

Fig. 6. Passenger Satisfaction on Environment and Health Friendliness

Figure 6 revealed that 81% of the commuters find the current tricycle system displeasing when it comes to environmental and health friendliness. The current tricycle system produces air pollution, which is for some unbearable and requires them to wear face masks. The tricycle system produces hefty amount of smoke emission which is unhealthy for both the commuter and the community it traverses.

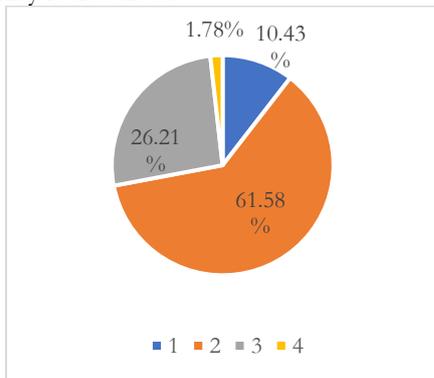

Fig. 7. Passenger Satisfaction on Facility

Figure 7 shows that 72% of the commuters find the facility of the current tricycle system displeasing. The current tricycle system has no legitimate terminal for passengers. Also, it underscores the need for a decent waiting area, deployment system and other support infrastructure needed in the tricycle system.

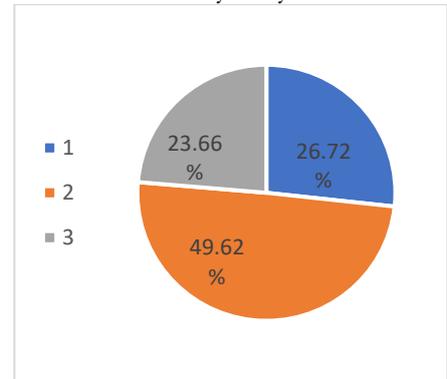

Fig. 8. Preferability on E-Trike

From Figure 8, the commuters are open on shifting to E-trikes where in 76% prefers to use the E-trike over the traditional trikes, while the remaining 24% could still opt it as their means of transport.

| Factors to Shift | 1 | 2 | 3 | 4 | 5 | 6 | Weighted Total | Ranking |
|---|---|---|---|---|---|---|---|---|
| Reliability | 21 | 18 | 29 | 19 | 23 | 22 | 467 | 3 |
| Comfort and Convenience | 21 | 34 | 24 | 20 | 16 | 16 | 417 | 6 |
| Safety and Security | 27 | 21 | 19 | 20 | 26 | 18 | 444 | 5 |
| Affordability | 21 | 17 | 17 | 26 | 18 | 32 | 492 | 2 |
| Environment and Health Friendliness | 19 | 20 | 15 | 21 | 31 | 25 | 493 | 1 |
| Facility | 22 | 21 | 26 | 25 | 19 | 18 | 445 | 4 |

Table 2. Drivers Demographics and Driving Profile

Table 2 shows that commuters prioritize environmental and health friendliness of the tricycle over everything else. The commuters perceived benefits in having an improved air quality that promotes the public health of the city's populace. Moreover, the commuters are willing to increase their transportation budget to increase each factor.

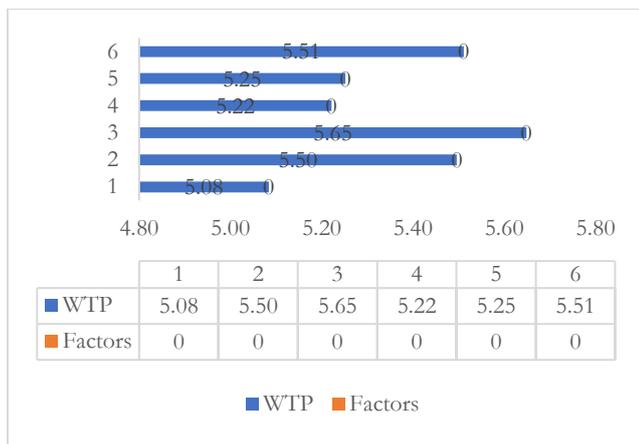

Figure 9. Summary of Commuters' Willingness to Pay for each Factor

Figure 9 shows that there are 16.10% commuters that are willing to pay additional PHP 5.65 on top of existing fare to improve their tricycle riding experience in terms of safety and security. Similarly, an additional PHP 5.51 is the willingness to pay when it comes to facility while 17.87% of the respondents agreed to add PHP 5.25 with the current fare by commuters to improve the environment aspect. Overall, an average of PHP 5.37 is the willingness to pay of the commuters in improving the current tricycle service in the city.

### 3.2 Demographics and Driving Profile

| A. Socio Demographic Attributes | |
|---|---|
| Sex | |
| Male | 100% |
| Female | 0% |
| Age | |
| <18 | 3.17% |
| 18-30 | 44.44% |
| 31-40 | 31.60% |
| 41-50 | 12.70% |
| 51-60 | 7.94% |
| >60 | 1.59% |
| B. Driving Pattern | |
| Average Number of Hours Driving (per day) | = 11.43 hours |
| Standard Deviation | = 7.87 |
| Average Daily Distance Travelled (in kilometer) | = 64.69 kilometers |
| Standard Deviation | = 62.50 |
| Gross Daily Income | |
| <PhP 200 | 7.94% |
| PhP200 – PhP400 | 28.57% |
| PhP400 – PhP600 | 33.33% |
| PhP600 – PhP800 | 23.81% |
| PhP 800 – PhP1000 | 6.35% |
| >PhP1000 | 0% |
| Average Daily Expenditures (gasoline, tariff, 1 meal) | = PhP111.59 |
| Standard Deviation | = 112.29 |

Table 3. Drivers Demographics and Driving Profile

Table 3 shows the demographic profile of drivers/respondents participated in the survey. All the respondents were male, and majority of them belongs to the 18-30 age bracket at 44.44%, followed by the 31-40 age bracket at 31.60%. I t can inferred that tricycle drivers' ages belongs to the working age of the population in the city. The driver's total number of operating tricycles in a day for a living is average at 11.43 hours with a standard deviation of 7.87. Moreover, the average daily gross income is recorded between Php 400 – Php600 which is 33.33% of the total respondents while Php200 – Php 400 followed with 28.57% response. The aggregate daily expenditure of a tricycle driver is Php112.59 which include fuel expenses, lunch meal, tariff and ticket, and parts of maintenance expenses.

The drivers then asked their knowledge about electric tricycle. The result shows that 88.89% of the respondents said they knew e-tricycle, while 11.11% replied "NO". Even though it is not yet been used in the city, drivers got this information about e-tricycle from neighboring cities specifically in Davao City where it is already implemented. Other drivers responded that they knew e-tricycle because they already had the opportunity to drive when they still stay with other places where there is operation of e-tricycle.

The survey revealed that drivers would really want to shift in using e-tricycle in their daily operation to earn a living given the favorable opportunity. 96.83% said they wanted to shift in using e-tricycle while 3.17% of the respondent said "NO".

Similarly, the drivers were asked about the time to be allotted in shifting to e-tricycle use. Majority of the respondents said that "1 – 2 years" (45.80%) is their time frame to shift to e-tricycle given the favorable opportunity while "3 – 4 years" garnered 31.75%. Drivers consider favorable opportunity as a timing for them to prepare a relatively low capitalization cash out from their end, and the improvement of the social environment amidst the pandemic, supporting their operation in the city and the creation of more demands for travel using e-tricycle.

| Perceived Problem | Frequency | Relative Weight | Rank |
|---|---|---|---|
| Money to buy E-tricycle | 352 | 5.59 | 1 |
| Difficulty in adopting to new technology | 301 | 4.78 | 2 |
| Political pressure | 286 | 4.54 | 3 |
| New requirement needed | 270 | 4.29 | 4 |
| Limited franchise route | 263 | 4.17 | 5 |
| Needed facility for E-tricycle operation | 252 | 4.00 | 6 |
| Others | 71 | 1.13 | 7 |

Table 4. Perceived Problems in Shifting to E-Tricycle

Table 4 shows that the principal concern of the drivers on shifting to e-tricycle is the needed capitalization cash out to acquire. Second in the rank is the difficulty in adopting to new technology. The respondents think that learning the operation of e-tricycle will take their time in mastering the skill. Other perceived problems include political pressure, new requirement needed, limited franchise route and support facility of e-tricycle operation.

It is noteworthy to highlights that although the category "Others" came last in the ranking, the respondent had provided some other problems like anxiety of driving and limited income.

| Needs | Frequency | Relative Weight | Rank |
|---|---|---|---|
| Financial help | 364 | 5.78 | 1 |
| Boundary-Pay system | 297 | 4.71 | 2 |
| Training of operating E-tricycle | 282 | 4.48 | 3 |
| Creation of drivers/operators cooperative | 269 | 4.27 | 4 |
| Better tricycle operation regulation | 262 | 4.16 | 5 |
| Simplified application of franchise | 255 | 4.05 | 6 |
| Others | 66 | 1.05 | 7 |

Table 5. Needs of Drivers/Operators to Shift to E-Tricycle

Table 5 revealed that the priority needs that the respondent identified to shift to e-tricycle will be financial help. Drivers perceived that organizations both from the government and private sector can be of an essential assistance to acquire a unit of e-tricycle. Second in the rank is the adaptation of boundary-pay system that will provide them lenient scheme in paying. Third in the ranking is the needs for training about e-tricycle. Other needs highlighted were the creation of cooperatives, better tricycle operation regulation and simplified application of franchise.

| Perceived Benefits | Frequency | Relative Weight | Rank |
|---|---|---|---|
| Better driving experience | 425 | 6.75 | 1 |
| Cleaner air | 344 | 5.46 | 2 |
| Higher income | 318 | 5.05 | 3 |
| Reduction of noise | 315 | 5.00 | 4 |
| Job opportunities/ additional livelihood | 304 | 4.83 | 5 |
| Improved management of tricycle operation | 274 | 4.35 | 6 |
| Better facilities | 258 | 4.10 | 7 |
| Others | 64 | 1.02 | 8 |

Table 6. Perceived Benefits of Shifting to E-Tricycle

Table 6 show that the respondents ranked of having a better driving experience as the most perceived benefits on shifting to e-tricycle with a relative weight of 6.75. Improvement in the air quality is also identified as an important benefit on having e-tricycle in the city. Moreover, drivers expect to obtain higher income in using e-tricycle for their daily operation. Other perceived benefits identified were reduction of noise, job opportunities, improved management of the tricycle operation and creation of better facilities as support infrastructure for e-tricycle operation.

| Hazards | Frequency | Relative Weight | Rank |
|---|---|---|---|
| Travel speed | 431 | 6.84 | 1 |
| Assembly of passenger's compartment | 335 | 5.32 | 2 |
| Physical configuration of the tricycle | 326 | 5.17 | 3 |
| Smoke emission | 317 | 5.03 | 4 |
| Road hazards | 301 | 4.78 | 5 |
| Design of passenger seats | 285 | 4.52 | 6 |
| Noise | 251 | 3.98 | 7 |
| Others | 64 | 1.02 | 8 |

Table 7. Identified Hazards of Existing Tricycle Operation

Table 7 explains that to further support of shifting to e-tricycle, the respondents were asked to identify the hazards of existing tricycle operating in the city. The respondents ranked travel speed as the primary hazard of the existing tricycle as it affects riding passengers thinking their safety of travel. Second in the ranking is the assembly of passenger's compartment. The current configuration mostly caters passenger that do not have disability. Limitation is also identified to the dimension that can be occupy by passengers once on board. Other hazards were smoke emission coming from the burning of fuel while the tricycle is moving, road hazards specially the poor visibility when it rains, passenger seat that sometimes accommodate more than what is the actual capacity, and noise induced.

### 3.3 Air Pollutants and Greenhouse Gas Emission from Motorized Tricycle

Emission from motorized tricycle greatly contribute to the increasing air pollutants that deteriorate the air quality which people breathe. Usual agent of various respiratory diseases that can be attributable to burning petroleum products such as Carbon Monoxide (CO), Hydrocarbon (HC), particulate matters (PM) and Nitrous Oxides (NOx). Similarly, the current trend of climate change impact to the increasing numbers of greenhouse gas emission, is attributable to petroleum products to power motorized vehicles. Notable GHG resulting from burning of petroleum products include $CO_2$, $CH_4$ and $N_2O$.

The actual visitation of emission testing centers in General Santos City and the obtaining of emission testing result had derived the following data that represents emission of air pollutants from motorized tricycle. Measurements of emission provide a baseline understanding of a given source compared to other sources and in initiating inventories of emission.

Based from the Intergovernmental Panel on Climate Change (IPCC) the emission factor of burning petroleum products as fuel for motorized vehicles is set at 2.27 kg $CO_2e$ for one (1) liter of gasoline, 2.63 kg. $CO_2e$ per liter of diesel. Using the mean daily distance travelled by tricycle at 64.69 kilometers to obtain the average number of liters of gasoline use.

IPCC Guidelines
Efficiency Rate: 7 kilometers per liter of gasoline
Emission factor: 2.27 kilograms $CO_2e$/ liter of gasoline
2.63 kilograms $CO_2e$/ liter of diesel

Applying the simple derivation to obtain the daily GHG emission ($CO_2e$)

GHG Emission = Activity data, (IPCC set) X Emission Factor
= 64.69 kilometers/ 7 kilometers/ liter of gasoline X EF
= 9.2414 liters of gasoline
= 9.2414 liters of gasoline X 2.27 kg $CO_2e$/ liter of gasoline
= 20.98 kilograms $CO_2e$ per tricycle per day

Referring to Asian Development Bank (ADB) Report that for the next decade, the massive increase in Carbon Dioxide emission due to the use of motorized vehicles including tricycle. The need for alternative mode that is cleaner and greener is very essential as the rate of $CO_2e$ emission alarms the possible negative impact to the people and to the environment. The initiative of ADB to introduce e-tricycle aim to provide a more sustainable transportation that will stir environmental protection such as improved air quality and reduced carbon dioxide emission.

### 3.4 Exploratory Data Analysis

Exploratory Data Analysis (EDA) is an approach/philosophy for data analysis that employs a variety of techniques (mostly graphical) to:
- maximize insight into a data set;
- uncover underlying structure;
- extract important variables;
- detect outliers and anomalies;
- test underlying assumptions;
- develop parsimonious models; and
- determine optimal factor settings.

EDA is a data analysis approach, that adapt the sequence
Problem => Data => Analysis => Model => Conclusions

For EDA, the data collection is not followed by a model imposition; rather it is followed immediately by analysis with a goal of inferring what model would be appropriate. Unlike the classical model approach, the Exploratory Data Analysis approach does not impose deterministic or probabilistic models on the data. On the contrary, the EDA approach allows the data to suggest admissible models that best fit the data. The focus of EDA is the treatment and existence of data – its structure, outliers and the models suggested by the data. The technique in EDA is generally graphical that includes scatter plots, character plots, box plots, histograms, probability plots, residual plots and mean plots. It differs from the classical technique as it relies heavily on quantitative presentations that include ANOVA, t test, chi-squared test, and F test. The advantage of employing EDA in analysis technique made use of the available data of the study that will provide little or no assumption – putting emphasis on the importance of presenting and showing the data as it is with fewer assumptions.

### 3.5 Python Programming Language

Python is an open-source programming language that is highly applicable with problems and researches involving statistical analysis. Halswanter (2015) considered Python as the most elegant programming language, adding that the programming language is free and powerful as far as statistical analysis is concerned.

Python has three different level of modularization that will enable result in an orderly manner. The levels of modularization are as follows:
- Function – defined by the keyword def, and can be defined anywhere in Python. Its return statement, typically at the end of the function;
- Module – is a file with the extension".py". It contains function and variable definitions, as well as valid Python statements; and
- Packages – is a folder which contains multiple Python module, normally bear a file named ___init___.ph.

Moreover, Halswanter (2015), had included the *statsmodels* which was developed by the Statsmodels Development Team (https://www.statsmodels.org/) had significantly been a part in Python package as a tool for modelling. *Statsmodels* enable to generate increased functionality of statistics for statistical data analysis, specifically provides classes and functions for the many different statistical models, as well as conducting statistical tests and statistical data exploration. *Statsmodels* also allows formulation of models on the notation introduced by Wilkinson and Rogers (1973). The commands od *statsmodels* have been applicable that it was tested against existing statistical packages ensuring the correctness of derived model. The data visualization as defined with *seaborn* is a Python visualization library based on matplotlib. It provides a concise high level interface for extracting statistical graphics that are both informative and attractive (Halswanter, 2015).

### 3.6 EDA Data Presentation using Python

As seen in Figure 10, it was observed that majority of the respondents have revealed their interest in shifting to E-tricycles from conventional motorized tricycles which means that when modelling based on these data the researcher must address the imbalanced class problem. For such an imbalanced class problem, the researcher used under sampling methods to try to balance the classes.

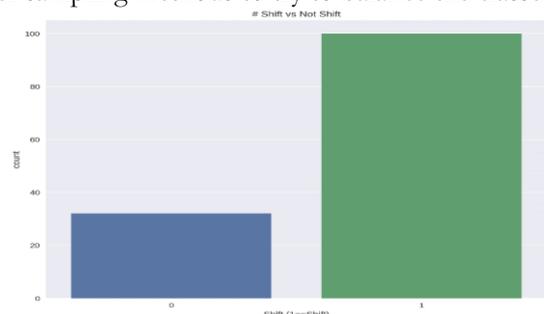

Fig. 10. Respondents Choice of Shift vs. Non Shift

To check multicollinearity, Figure 11 shows the correlation matrix of all the predictors (Environment, Safety, Reliability, Comfort, Facility, Affordability). There is no significant correlation among the predictors. Hence, dimensionality reduction is not necessary.

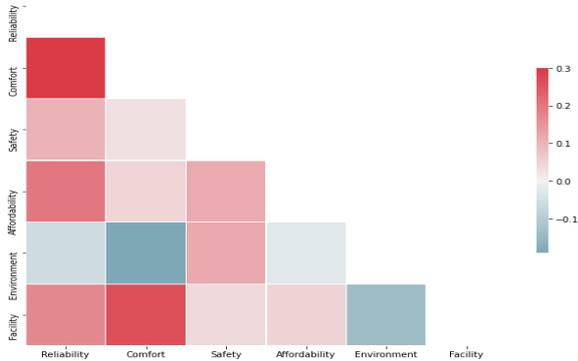

Fig. 11. Correlation Matrix of Predictors

Binary Logistic Regression

- To model the probability of e-tricycle shift, the researcher used Environment, Safety, Reliability, Comfort, Facility and Affordability scores as predictors.
- In Logistic Regression the logits are assumed to be a linear function of the predictors.

$$L = \log\left(\frac{P(Y=1)}{1-P(Y=1)}\right) = \beta_0 + \sum_{i=1}^{n} \beta_i X_i \quad (1)$$

- To solve for P(Y=1),

$$p = \frac{e^L}{1-e^L} \quad (2)$$

- The parameters βi can be derived by Maximum Likelihood Estimation (MLE). The likelihood for a given *m* observation *Yj* is;

$$lkl = \prod_{j=1}^{m} p^{Y_f}(1-p)^{1-Y_j} \quad (3)$$

- Training the model: The dataset is split into training set and test set (70% of the samples are used for training and 30% is used to test the model.

$$L = 1.15984927 + 0.05992501 X_{Reliability}$$
$$- 0.42234063 X_{comfort} + 0.37066915 X_{safety}$$
$$- 0.1853214 X_{affordability} + 0.41213324 X_{environment} + 0.10333073 X_{facility}$$

The derivation of logistic regression model that the predictors of a shift to e-tricycle as mode of new transport in General Santos City is best explained that the Reliability, Safety, Environment and Facility predictors were factors that will support the shift as derived with their positive coefficients. On the other hand, predictors Comfort and Affordability suggested its negative coefficients to be of less consideration based from the equation model. To further fine tune the derived equation model, it is essential to evaluate the model that will represent the data being treated using EDA.

➤ Model Evaluation: Based on the confusion matrix shown in figure 4.11, the derived model was able to label 24 true positives (correctly predicted shift) out of 33 samples from our test dataset which yields 72.72% model accuracy.

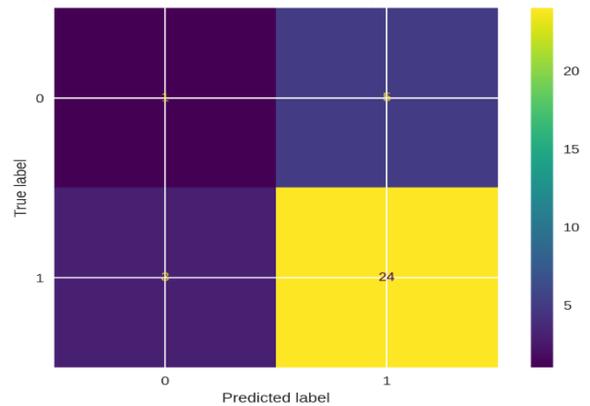

Fig. 12. Confusion Matrix

➤ However, accuracy is not enough to evaluate the model since the data suffers the *imbalanced class problem*. Hence, we should also look at the AUC-ROC curve. The Receiver Operator Characteristic (ROC) curve is an evaluation metric for binary classification problems. It is a probability curve that plots the TPR against FPR at various threshold values and essentially separates the 'signal' from the 'noise'.

➤ The Area Under the Curve (AUC) is the measure of the ability of a classifier to distinguish between classes and is used as a summary of the ROC curve. The higher the AUC, the better the performance of the model at distinguishing between the positive and negative classes. For this model the AUC is 72.47 %

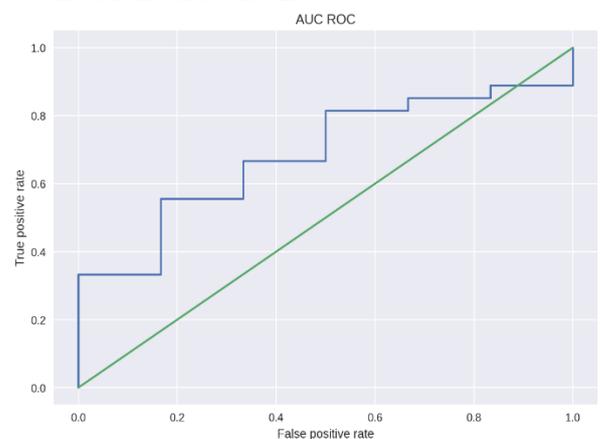

Fig. 13. Area Under Curve Graph

The algorithm of Phyton programming analysis of the data set is attached to as appendix A showing all the inputted syntax for the analysis and model equation derivation.

## 4. Conclusion and Recommendation

The adoption of e-tricycle in General Santos City is an after sought alternative that will provide sustainable means of moving local residents of the city and improve their lives. As e-tricycle promotes cleaner technology means improving the air quality of the entire city, it also advances safe and comfortable means of physical movement amongst passengers. Referring to the "Three E's" of sustainable transportation, e-tricycle can provide:

Environment: better air quality since e-tricycle will not have emission of both greenhouse gases and air pollutants;

Equity: e-tricycle is anticipated to improve the social being of the city with the correlation of clean surroundings to improved quality of life;

Economy: shifting to e-tricycle will provide a scale of economy that will spur to the city as it will provide greater opportunity for both job and livelihood generation, and increased income for the drivers.

The following were sought out in this study:
- Based on the survey results, environment and health friendliness and reliability are the most important factors in shifting from traditional tricycles to e-trikes.
- Most of the respondents are not satisfied on their current riding experience due to smoke emitted and noise produced from the traditional tricycles.
- If aspects of the riding experience are improved, more than 50% are preferably choosing e-trikes.
- 80% of the respondents are willing to pay additional fare for the improvement of the tricycle service considering the six factors.
- The drivers' response revealed that they have knowledge about e-tricycle with its operation, maintenance and the perceived benefits once they will shift.
- The need for financial assistance to capitalize the acquisition of a unit of e-tricycle seems so imperative to the drivers as they see it as burden as far as availability of money is concerned.

- Hazards of the existing tricycle were identified to include travel speed and road hazards, the current configuration of the passenger's seat and compartment, smoke and noise emission among others.
- Given the opportunity of available resources, the drivers were willing to shift to e-tricycle for better income and services to their passengers.
- The average daily Carbon Dioxide emission of a motorized tricycle is 20.98 kilograms per unit per day. The obtained value signifies important contribution to the increasing number of $CO_2$ emission that will hasten the impact of climate change whether on local or global scale.

The exploratory data analysis in Python is really a dynamic tool that provided the model that explained the possibility of shift to e-tricycle. The model equation derived indicates that "reliability" and "environment" factors play significant role in the decision of the commuters to shift which provided a 72.72% accuracy using binary logit. It can be seen that the commuters of General Santos City are open for the shifting of the conventional tricycles to e-trike, given certain aspects that e-trike can offer in terms of ridership comfort. Consistent with the objective of CHED-PCARI DARE Project, the derived model of the study can be a basis for the local government to consider the adoption of e-trike in General Santos City that can be included in their public transportation policy. Similarly, the City Government of General Santos may include E-trike Project in their Comprehensive Development Plan (CDP) and Local Climate Change Action Plan (LCCAP). In the larger scale, The SOCKSARGEN Region could consider e-trike as their project proposal to access the People's Survival Fund and other international funding related to climate change. Climate change mitigation initiatives are more effective if implemented by regional level.

Moreover, the commuters are willing to allot an average additional PHP 5.37 to improve the current tricycle system. With comfort and convenience ranking as the highest contribution as a factor for commuters to pay additional fare and safety, security and reliability as the least. The most considered aspect of commuters in shifting to e-trike is environment and health friendliness and reliability. They are concerned in reducing the smoke emissions, noise and the timeliness of the tricycle when needed.

Further recommendation of the study, future researchers may include the consideration of the technical aspects of the e-trike (e.g. carriage structure, air conditioning, etc.) to make the basis of ridership comfort more accurate. Also, the consideration of willingness of passengers to use the e-trike in inclement weather conditions may be of help to determine the perception of commuters in this case. Thirdly, an initiative from the Local Government of General Santos City to conduct public consultation on adoption of e-tricycle that will focus on the possible schedule of shift to e-tricycle of all stakeholders. Lastly, the study strongly indorses the optimization on data gathering that will provide significant input in data analytics for the improvement of General Santos City's transportation system.

**Acknowledgement**

The authors would like to acknowledge the officials of the local government unit of General Santos City headed by Honorable Mayor Ronnel C. Rivera and his city officials Engr. Riza Paches, Engr. Myles Gecosala and Mr. Harold Aponesto.

**References**

[1] Asian Development Bank (2010) E-Trike Project: Clean Development Mechanism Credits.

[2] Asian Development Bank (2011) Implementation of Asian City Transport –Promoting Sustainable Urban Transport in Asia Project. *Sustainable Urban Transport Study in Davao City.*

[3] Bachok, Syahriah, Ponrahono, Zakiah, Osman, Mariana Mohamed, Jaafar, Samsuddin, Ibrahim,Mansor, Mohamed, MohdZin (2015). The 5th Sustainable Future for Human Security. *A Preliminary Study of Sustainable Transport Indicators in Malaysia: The Case Study of Klang Valley Public Transport,* doi: 10.1016/j.proenv.2015.07.056 pp.464-473.

[4] Balaria, F., Pascual, M., Santos,M., Ortiz, A., Gabriel, A., Luz, T., Mangahas, S. (2017). Open Journal of Civil Engineering. *Sustainability of E-Trike as Alternative Mode of Public Transportation System: The Case of Cabanatuan City.* pp 362-377

[5] Benetto, Barabino, Eusebio, Deiana (2012). SIDT Scientific Seminar 2012. *On the Attribute And Influencing Factors of End-Users Quality Perceptions in Urban Transport: An Exploratory Analysis.* doi:10.1016/j.sbspro.2013.10.591 pp 18-30.

[6] Camagay, M. D., Sigua, R. G., Vergel, K. N., Matias, A. C., & Danao, L. M. (2005). Journal of the Eastern Asia Society for Transportation Studies. *Development of Emission and Engine Testing Procedures and Standard Sidecar Design Prototype for Tricycles,* 6, 3151-3166.

[7] Capricho, Lourdes Maria A. (2018) Energy Consumers and Stakeholders Conference. *Alternative Fuels Vehicle and Energy Technology E-Power Mo!.*

[8] Commission on Higher Education – Philippine-California Advance Research Institute: *Data Analytics for Research and Education* (CHED-PCARI DARE), November 2017 – October 2019.

[9]Dakis, K. J., Chavez, S. M., & Bandril, R. (2018). *An Assessment of the Pilot Implementation of the Tricycle Modernization Program of the City Goverment of Imus, Cavite Province.*

[10] Declan, D. N. (2012). Journal of Social Sciences and Public Affairs Vol 2 No. 2. *An Empirical Study of the Use of Tricycle as a Public Transport Mode in Nigerian Cities.* pp 66-76.

[11] Department of Energy, Republic of the Philippines (2010). *Market Transformation through Introduction of Energy Efficient Electric Vehicle Project Document.*

[12] General Santos City, Republic of the Philippines (2017). *Local Public Transportation Route Plan, Final Report.*

[13] Hall, Ralph P., (2002). Introducing the Concept of Sustainable Transportation to the U.S. DOT through the Reauthorization of TEA-21. Graduate Thesis, Massachusetts Institute of Technology, June 2002

[14] Halswanter, Thomas (2015) An Introduction to Statistics with Python with Application in the Life Science. Springer International Publishing AG Switzerland DOI 10.1007/978-3-319-28316-6

[15] Kumar, Gaurav, Kaur Amandeep, Singh, Kiran K., (2014). International Research Journal of Social Science. *Public Transport and Urban Mobility: Perception of People on Services of Public Transport in Bathinda City, Punjab, India.*

[16] Litman, Todd (2012). The Annual Review of Public Health. *Transportation and Public Health.* Annual Review Public Health 2013, 34.22.1-22.17 Available online: www.publichealth.annualreviews.org, doi:10.1146/annurev-publichealth-031912-114502.

[17] Luansing, Ronalyn, Pesigan, Clarissa, Rustico Jr., Eduardo, (2015) 6th International Conference on Applied Human Factors and Ergonomics 2015. *An E-trike Project - Innovative, Concrete and Ergonomic: Systems Design to Support Sustainable E-Trike Commercialization.* AHFE Conference, doi:10.1016/j.promfg.2015.07.380. 2833-2840.

[18] Melieres, Marie-Antoinette, Marechal, Chloe (2015). *Climate Change: Past, Present and Future 1st ed.* Published by John Wiley & Sons, Ltd. Companion website: www.wiley.com/go/melieres/climatechange

[19] Melo, Sandra, Baptista, Patricia, Costa, Alvaro (2013). *Comparing the Use of Small Sized Electric Vehicles with Diesel Vans on City Logistics.* Procedia - Social and Behavioral Sciences 111 (2014) pp 350-359 published by Elsevier Ltd. doi:10.1016/j.sbspro.2014.01.068

[20] Mohan, Dinesh, Tiwari, Geetam (1999) *Sustainable Transport System: Linkages Between Environmental Issues, Public Transport, Non-Motorised Transport and Safety.*

Prepared for the Wuppertal Institute, Germany. Reprinted from Economic and
Political Weekly, Vol XXX14:25, 1999, 1589-1596. Posted at www.vtpi.org with author's permission.

[21] NIST/SEMATECH e-Handbook of Statistical Methods,
https://www.itl.nist.gov/div898handbook, 2012. DOI 10.18434/M32189

[22] Pietzrak, Krystian, Pietzrak, Oliwia, (2020) Sustainability 2020. *Environmental Effects of Electromobility in a Sustainable Urban Public Transport. Sustainability 2020, 12,*
1052; doi:10.3390/su12031052. Available online:
https://www.mdpi.com/journal/sustainability

[23] Sarsalejo, Lady Fritz C., Preclados, Lemuel S. (2018) Journal of Education and Human
Resource Development 6:1-11. Comparative Profitability Analysis of Electric,
Pedicab, and Gasoline-Fuelled Tricycles. ISSN 2545-9759.

[24] Sumaedi, Sik, Bakti,Gehe Mahatma Yuda, Yermen, Medi (2012) International Journal for
Traffic and Transport Engineering. *The Empirical Study of Public Transport
Passengers' Behavioral Intentions: The Role of Service Quality, Perceived Sacrifice,
Perceived Value, and Satisfaction: Case Study: Paratransit Passengers in Jakarta,
Indonesia,* pp. 83-97.

[25] Washington, Simon P., Karlaftis, Matthew G., Mannering, Fred L., (2020). *Statistical and
Econometric Methods for Transportation Data Analysis, 2nd edition.* pp.283-372

[26] Watson, R.T., Zinyowera, M.C., Moss, Richard H. *The Regional Impacts of Climate Change,
An Assessment of Vulnerability*. Intergovernmental Panel on Climate Change (1998)
*Cambridge University Press, United Kingdom.*

[27] Woldeamanuel, Mintesnot G., Cyganski, Rita (2011). Association for European Transport and
Contributors 2011. *Factors Affecting Travellers' Satisfaction with Accessibility to
Public Transportation.*

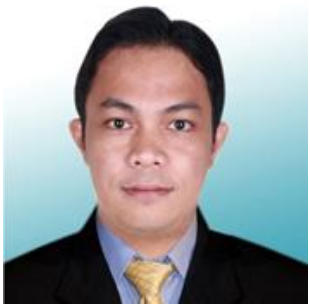
**G.L. Cueto** is a Master of Science student at Mapua Unviersity. He finished is BS Civil Engineering at New Era University in 1995, Diploma in Transportation Planning at University of the Philippines Diliman in 2002 and BS in Sanitary Engineering at National University in 2018.  His major field of study includes public transportation, traffic impact assessment and traffic engineering.

   He is currently working as Project Director in a construction firm providing professional service on project managment, program and system assessment and audit, traffic impact assement and traffic management plan and occupational safety and health.  Also, he is currently a faculty member of the School of Civil, Environmental and Geological Engineering of Mapua University where he handles undergraduate courses on Traffic and Transportation Engineering, Traffic Impact Assessment and Ports and Harbor Engineering.

   Engr. Cueto is an active member of the following professional organization to include Philippine Institute of Civil Engineers (PICE), Philippine Institute of Environmental Planners (PIEP), Philippine Society of Sanitary Engineers (PSSE), Transportation Science Society of the Philippines (TSSP) and International Water Association (IWA).  He is a resident lecturer of seminars on environmental seminars for Pollution Control Officers.

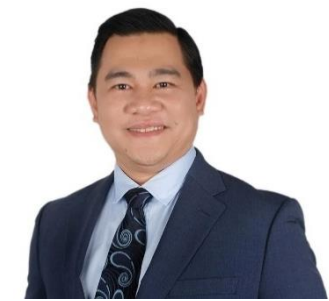
**F.A.A Uy**  is the Founder and President USHER Technologies Inc.  Dean, School of Civil, Environmental and Geological Engineering of Mapua University, Philippines  Dr. Francis Aldrine Uy is the founder and president of USHER Technologies, the first Mapua spin-off company derived from the Universal Structural Health Evaluation and Recording System (USHER) project of DOST-PCIEERD and Mapua University.  The USHER System is a 24/7 structural health monitoring online platform for infrastructures that greatly improves our preparation and response to strong earthquakes and typhoons. Dr. Uy is the DOST-PCIEERD 2018 Outstanding R&D Awardee on Special Concerns and 2018 World Summit Awards Winner for Smart Settlements & Urbanization Category. Since the year 2008, he is the Dean of the School of Civil, Environmental and Geological Engineering at Mapua University.  He took his Doctor of Philosophy in Civil

Engineering at the University of the Philippines Diliman, Master of Science in Civil Engineering at the Technological University of the Philippines and Bachelor of Science in Civil Engineering at the Mapua Institute of Technology. He is a graduate of Leadership in Innovation Fellowship (LIF 4 Fellow) at the Asian Institute of Management under the School of Executive Education. LIF is part of the Newton Agham Programme, a collaboration between the UK and the Philippine government, the UK Royal Academy of Engineering, DOST, and AIM. He is one of The Outstanding Mapuans (TOM) 2015 awardee in the field of Academe given by the National Association of Mapua Alumni (NAMA). He is a recipient of the 2019 Outstanding ASEAN Science Diplomat Award. Dr. Uy and the USHER system was also internationally recognized in 2019 with the Outstanding Engineering Achievement Award given by the ASEAN Federation of Engineering Organizations (AFEO). He is one of the 2019 Manila Water Foundation Engineering Excellence Prize awardee. He was awarded with the David M. Consunji Award for Engineering Research 2020 given by the Philippine Association for the Advancement of Science and Technology (PhilAAST). He was awarded with the Kabalikat Researcher Award in line with DOST-PCIEERD's 10th year anniversary in 2020. Just recently because of USHER's fight COVID-19 initiatives, Dr. Uy has been chosen as one of the awardees of the 2020 Ginebra Ako Awards.

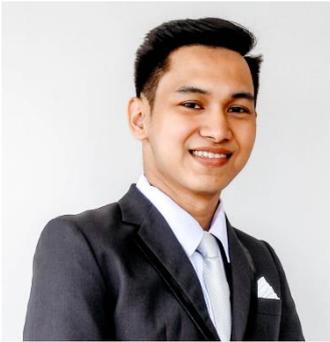

**K.A. Diaz** finished his B.S. and M.S. in Civil Engineering degrees from Mapua University, Manila in 2017 and 2020, respectively. During his graduate program, he joined the Mobile Sensing Lab at the University of California, Berkeley as a visiting student researcher. Keith is currently a faculty member of the School of Civil, Environmental and Geological Engineering of Mapua University where he handles Traffic and Transportation Engineering undergraduate courses. His research interests include intelligent transportation systems specifically the integration of machine learning and traffic engineering